\documentclass[sigconf]{acmart}
\usepackage{amsmath}
\usepackage{algorithm}
\usepackage[noend]{algpseudocode}

\usepackage{comment}

\def\BibTeX{{\rm B\kern-.05em{\sc i\kern-.025em b}\kern-.08emT\kern-.1667em\lower.7ex\hbox{E}\kern-.125emX}}
    
\copyrightyear{2018}
\acmYear{2018}
\setcopyright{acmlicensed}
\acmConference[Woodstock '18]{Woodstock '18: ACM Symposium on Neural Gaze Detection}{June 03--05, 2018}{Woodstock, NY}
\acmBooktitle{Woodstock '18: ACM Symposium on Neural Gaze Detection, June 03--05, 2018, Woodstock, NY}
\acmPrice{15.00}
\acmDOI{10.1145/1122445.1122456}
\acmISBN{978-1-4503-9999-9/18/06}

\begin{document}

%
\title{Optimizations to the Parallel Breath First Search on Distributed Memory}

%


\author{Anuj Sharma}
\affiliation{%
  \institution{Amazon, Inc}
  \streetaddress{1 Th{\o}rv{\"a}ld Circle}
  \city{Seattle}
  \country{United States of America}}
\email{amanuj8330@gmail.com}

\author{Syed Mohammed Arshad Zaidi}
\affiliation{%
  \institution{University at Buffalo, SUNY}
  \city{Buffalo}
  \country{United States of America}}
  \email{szaidi2@buffalo.edu}

%
\renewcommand{\shortauthors}{Sharma and Zaidi}

%
\begin{abstract}
Graphs and their traversal is becoming significant as it is
applicable to various areas of mathematics, science and
technology. Various problems in fields as varied as
biochemistry (genomics), electrical engineering
(communication networks), computer science (algorithms and
computation) can be modeled as Graph problems. Real world
scenarios including communities their interconnections and
related properties can be studied using graphs.
So fast, scalable, low-cost execution of parallel graph
algorithms is very important. In this implementation of
parallel breadth first search of graphs, we implemented Parallel BFS algorithm with 1-D partitioning of graph as described in \cite{bulucc2011parallel} and have reduced execution time by optimizing communication for local buffers.

\end{abstract}

%
%

%
\keywords{Graphs, Parallel Algorithms, Distributed systems}

%

%
\maketitle

\section{Introduction}
Fast and scalable graph algorithms are critical for a large
range of application domains with a vital impact on both
national security and national economy which includes social
media, logistics, e-commerce etc. Parallel scalable graph
algorithms are challenging in scenarios where we have
distributed memory architecture, as message passing includes
overhead of message passing between multiple nodes. Several studies has been recently done on BFS for parallel systems \cite{ueno2013parallel,chhugani2012fast,leiserson2010work}.
We got inspired for this work from the fact that large and
often scale-free, graphs are ubiquitous in communication
networks and social networks like that of Facebook. For
traversal on different types of graphs we used start graph,
Erd\H{o}s-R\'enyi graph \cite{blumofe1996dag} and small world graph. Through our results by showing success of this parallel BSF algorithm we hope to show potential of this proposed algorithm to wide range of parallel graph problems. We tried our level best to
bring out any limitations of this algorithm on certain types of
graphs wherever possible.

\section{Shared Memory Implementation}
Serial algorithms work on single machine with large enough
computing resources including memory and processor power.
Number of such serial implementations are already available.
Various challenges included in implementation of parallel
algorithm not only involved partitioning and distribution of
vertices among processors but also communication among all
processors, updating processor local data with others and
keeping global data update for next iteration of algorithm. It
also involved collecting, merging and distributing local data buffers in efficient way to reduce communication time.\ Various approaches used in tackling such challenges can be seen in depth further.

\subsection{Partitioning}
We used 1-D partitioning which partitions graph in following
steps:
\begin{enumerate}
\item Choose start vertex assign it to first processor.
\item Calculate all neighbors of start vertex.
\item In next step divide all n neighbors of start vertex among
$p$ processors $(n/p)$.
\end{enumerate}
In this way after completion of each phase we first gather
computed neighbors from all processors and partition them at
the beginning of next step.


\begin{figure}[h]
  \centering
  \includegraphics[width=\linewidth]{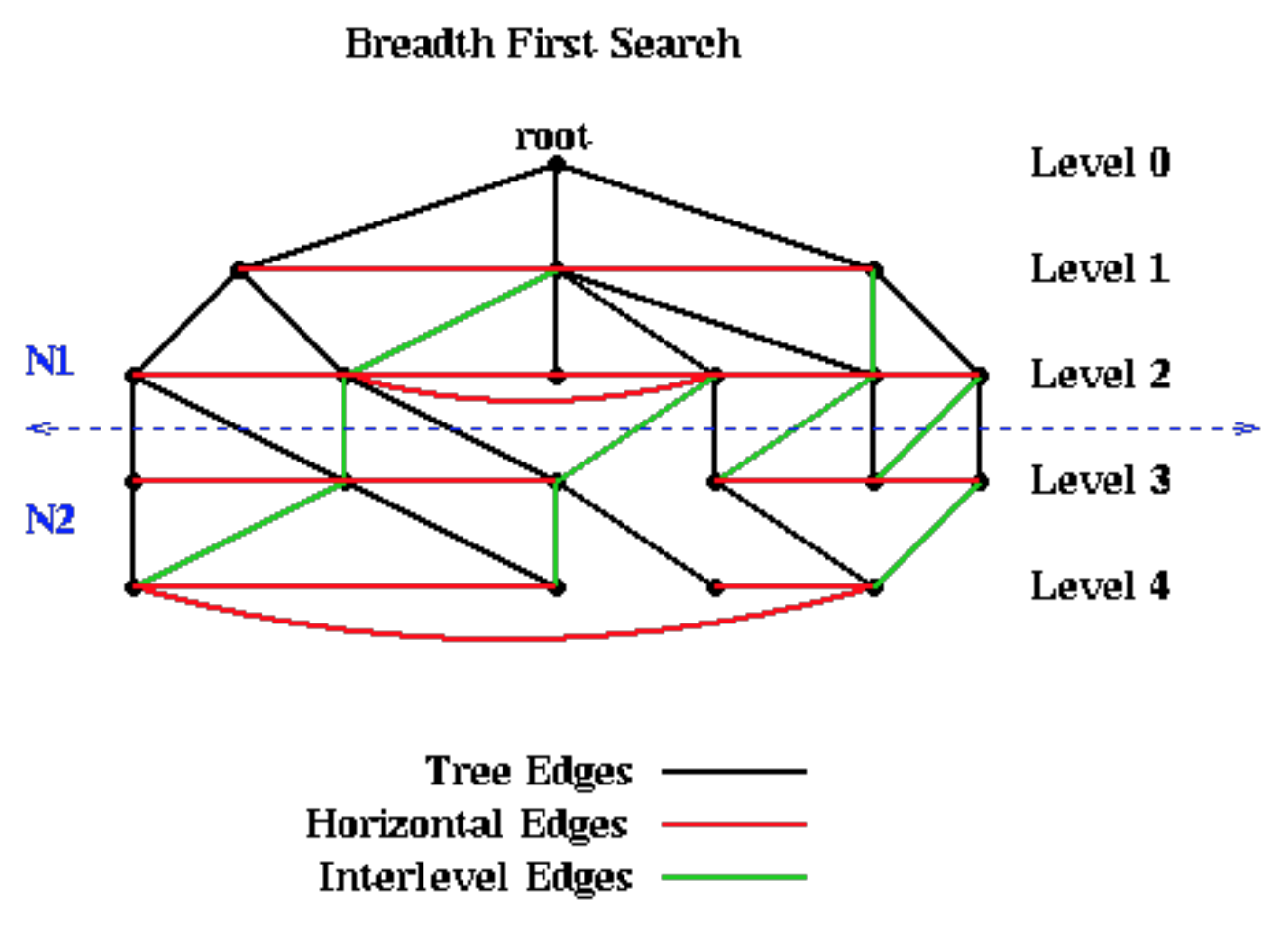}
  \caption{Breadth First Search}
  \Description{}
\end{figure}

In Figure 1, we can see that in step 1 Level 0 only processor 1 will have root vertex. In next step Level 1 there are three vertices and processor 1, 2 and 3 each will have one vertex. In next step Level 2 there are 6 vertices and each processor from 1-6 will have one vertex.
In this way partitioning of vertices will keep on happening till
all vertices are traversed.

\subsection{BFS Traversal}
Traversal begins with single start vertex and keeps on
incrementing with each step of finding all neighbors of vertex
in current hand for each processor.\ We have used the
following algorithm \cite{bulucc2011parallel} for the purpose of BFS traversal.

\begin{figure}[h]
  \centering
  \includegraphics[width=\linewidth]{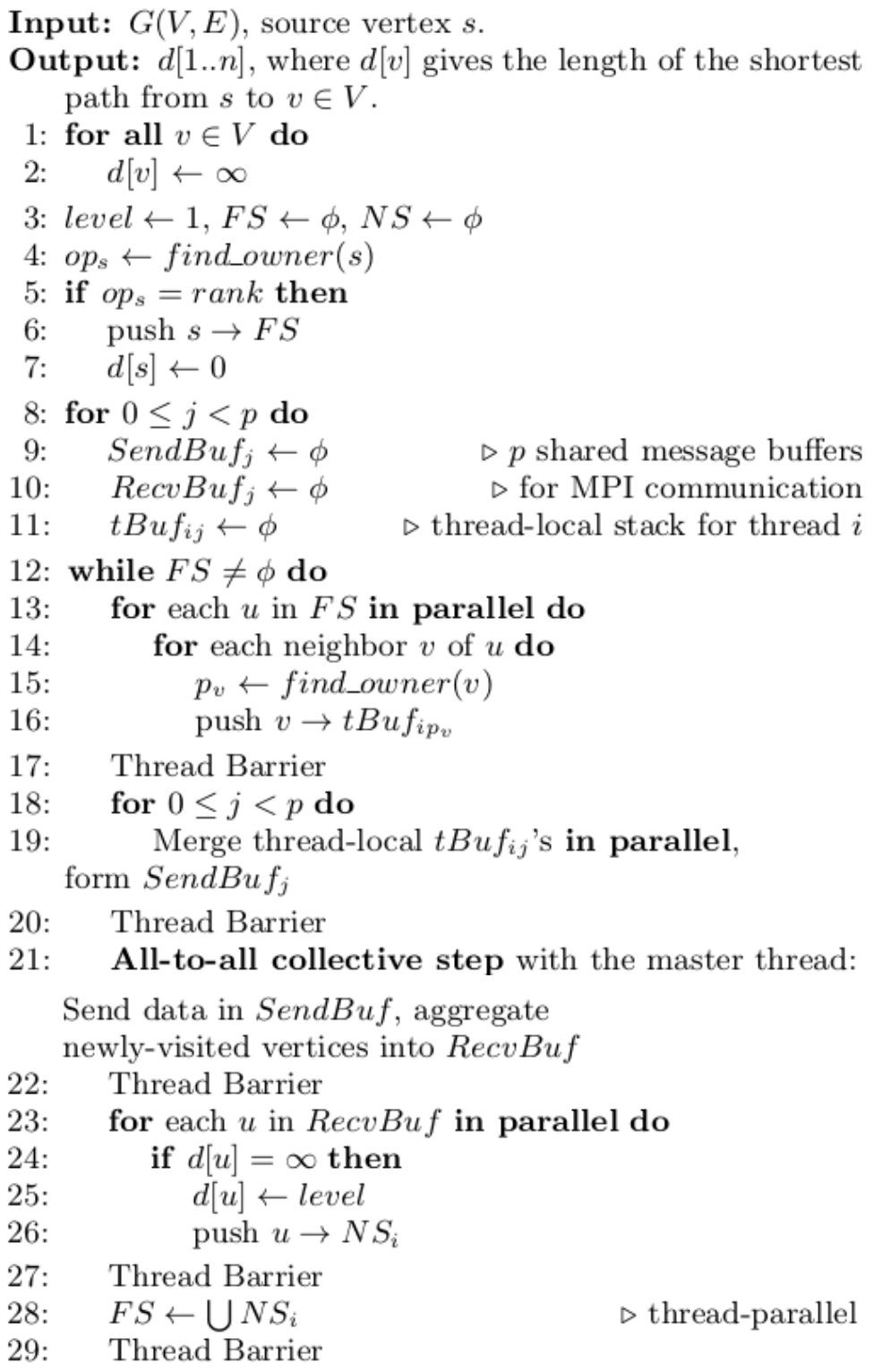}
  \caption{Algorithm for BFS traversal}
  \Description{}
\end{figure}

\makeatletter
\def\BState{\State\hskip-\ALG@thistlm}
\makeatother

\subsection{Global Computation and Communications}
Generalizing the whole algorithm comprises of two steps.
In the first step, the computation performed by each processor while in the second step, the communication is performed by all processors at
the end of each iteration.

For explaining computation step say we are in $n^{th}$ iteration
step, where total number of vertices are 8000 and we have
implemented on 8 processors. So according to 1-D partitioning
all these vertices will be partitioned in $(n/p)$ small chunks
among each processor. Now according to our example each
processor will have 1000 vertices. Then each processor will
perform local computation on each vertex finding all the
neighbors of this vertex. Only owner i.e. processor can decide
if vertex is visited or not and can assign level to it. Each
processor stores the same number of local buffers as of total
processors participating into BFS. In case neighbor found is
such that computing processor is the owner of such neighbor
then immediately processor will update its distance vector for that vertex.

After completion of computations by all processors, next
step involves communication and interchanging of data
between them. Every processor sends all of the local buffers
except the one it owns to other owner processors. Once
communication step is completed, each processor receives all
the vertices it owns, such vertices are evaluated again to
update their distance vector and then move on to next
iteration.

\section{Evaluation}
Basic results of scalability for a parallel code running on
parallel cluster system are strong and weak scaling. As strong
scaling has a constant problem size and increasing processor
or core count, it measures how well a parallel code can solve a
fixed size problem as the size of parallel computer is
increased. Whereas weak scaling has a fixed problem size per
core per processor and measures the ability of parallel
machine with parallel code to solve large versions of the same
problem.
To evaluate our approach, we conducted empirical study of
the strong and weak scaling on HPC cluster up to 64 nodes
involving combinations of processors and cores. we tested on
graphs consist of up to 4 Million vertexes. To check for any
anomaly, we checked our data for following types of graphs:
\begin{itemize}
\item Star Graphs
\item Erd\H{o}s-R\'enyi Graphs
\item Small World Graphs
\end{itemize}
Environment used in the implementation is described below
which precedes results.

\subsection{Test Environment}
Test environment is CPU cluster hosted at University at
Buffalo. For our algorithm execution we used from 1 to 64
nodes of cluster. While relying on CPU memory we faced
certain problems like while generation of graph with 4 million
vertices with distributed memory architecture we were running
out of memory again and again. To tackle such problems, we
divided set of work into small chunks of work. For example,
graph is generated for 1000000 vertices and then concatenated
locally to get resulting graph of 4 million vertices.\ Runtime
environment used was Open MPI which keeps tracks of all
participating processors and provides us with very handy
constructs like \textit{MPI\_Bcast} (for broadcasting), \textit{MPI\_Isend}
(Non-blocking send), \textit{MPI\_Recv} (for receiving data from any
rank), \textit{MPI\_Gatherv} (for uniting data from various
ranks/processors). For graph generation we used BOOST
library with C++ language. This library is perfect for
generating large graphs of different types.
While computations all data is stored at CPU memory and
all computations, for example, computing of all neighbors of
any vertex has been done on CPUs. On provided HPC cluster
we submitted jobs to SLURM job scheduler which optimizes
job scheduling and further prioritize jobs based on their
resources usage and priority.

\section{Results}
Scalability results of our algorithm demonstrated the
underlying promise of our approach. They basically contrasted
computation costs vs communication costs in some cases very
distinctively.

We implemented our algorithm on the following graphs:
\begin{itemize}
\item Star Graphs
\item Erd\H{o}s-R\'enyi Graphs
\item Small World Graphs
\end{itemize}

\subsection{Scalability Results with Star Graphs}

Parameters defined for star graphs are as follows:

\begin{figure}[h]
  \centering
  \includegraphics[width=\linewidth]{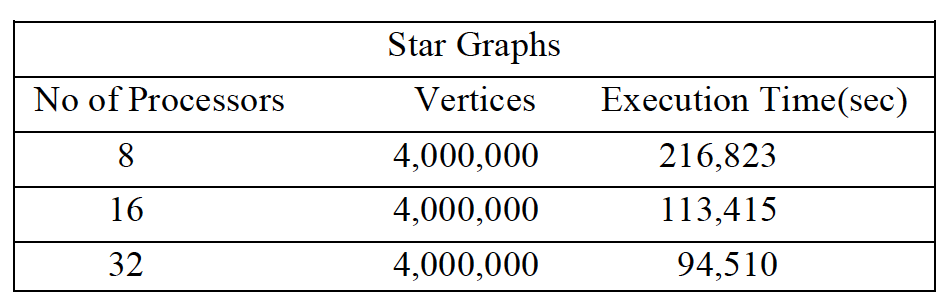}
  \caption{Parameters for star graphs}
  \Description{}
\end{figure}

\begin{figure}[h]
  \centering
  \includegraphics[width=\linewidth]{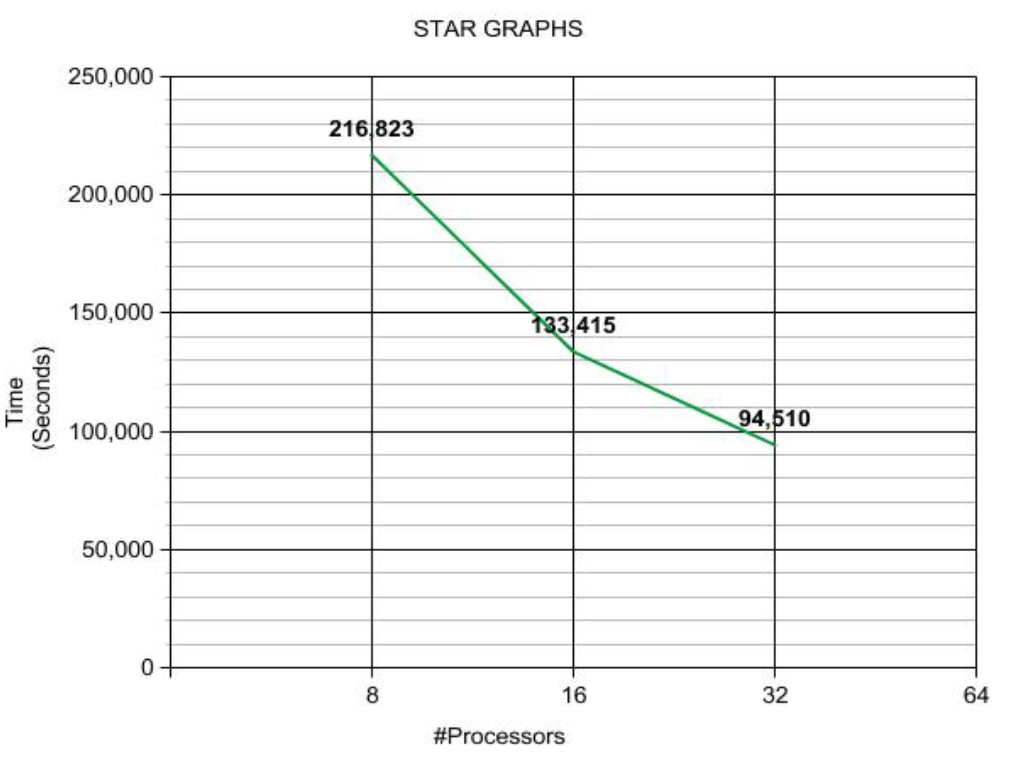}
  \caption{Scalability Results for Star graphs}
  \Description{}
\end{figure}

We were able to achieve these three computations for
4,000,000 vertices and processors 8, 16 and 32. For nodes like
1, 2, and 4 as stated in next sub-section CCR discarded jobs
due to time limit of 72 hours.
It showed strong scaling as we kept problem size fixed and
it showed decrease in execution time as we increased number
of processors.

\subsection{Scalability Results with Erd\H{o}s-R\'enyi}

\begin{figure}[h]
  \centering
  \includegraphics[width=\linewidth]{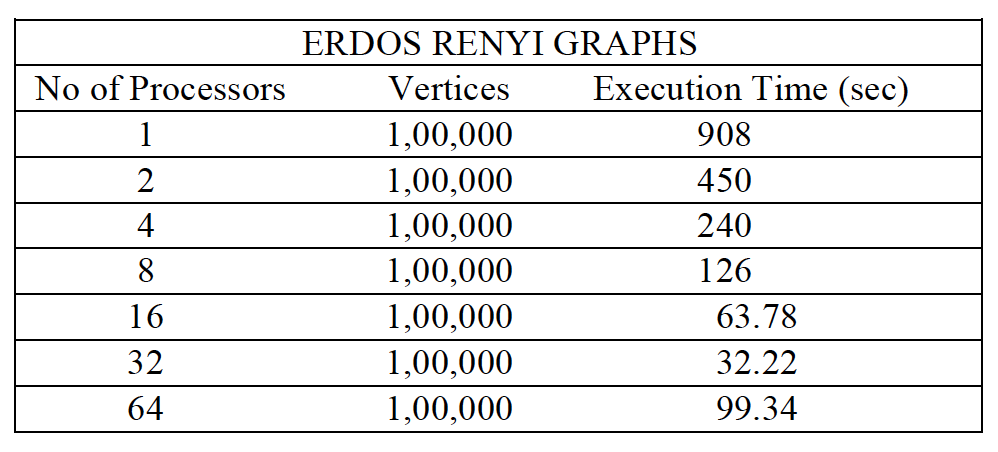}
  \caption{Parameters for Erd\H{o}s-R\'enyi graphs}
  \Description{}
\end{figure}

\begin{figure}[h]
  \centering
  \includegraphics[width=\linewidth]{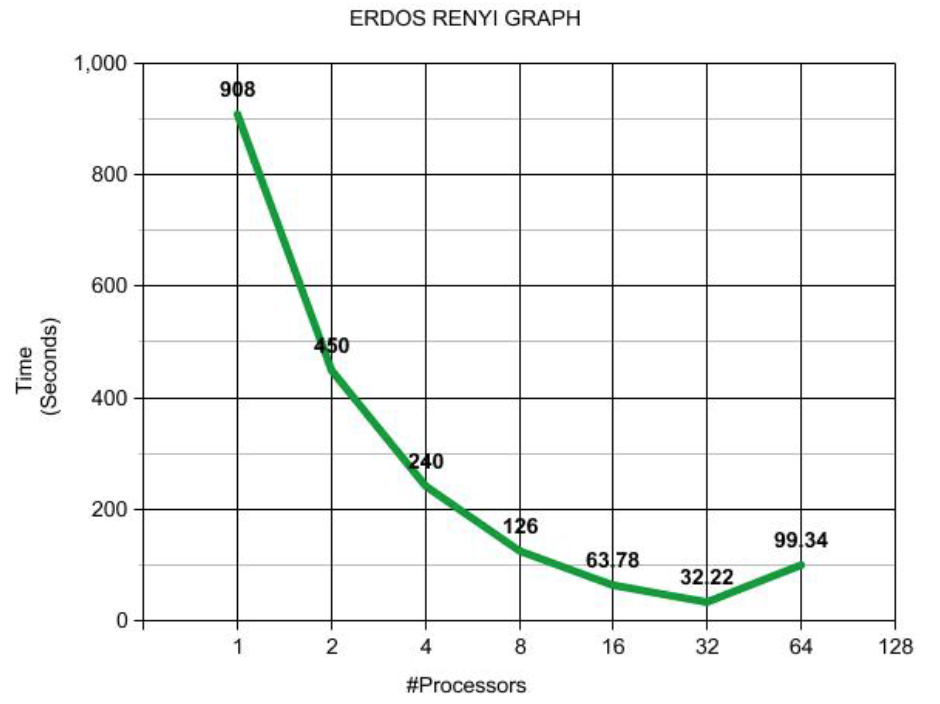}
  \caption{Scalability Results for Erd\H{o}s-R\'enyi graphs}
  \Description{}
\end{figure}

Results for Erdos Renyi graphs shows very nice scalability till
32 processors. After 32 processors when we checked it for 64
processors it increased which is caused by domination of
message passing overhead between processors over
computation time.

\subsection{Scalibility Results with Small World Graphs}

\begin{figure}[h]
  \centering
  \includegraphics[width=\linewidth]{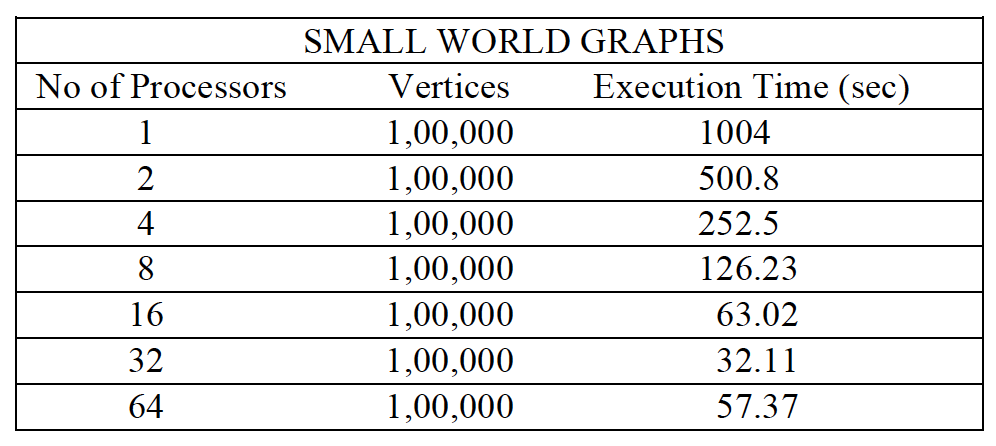}
  \caption{Parameters for Small world graphs}
  \Description{}
\end{figure}

\begin{figure}[h]
  \centering
  \includegraphics[width=\linewidth]{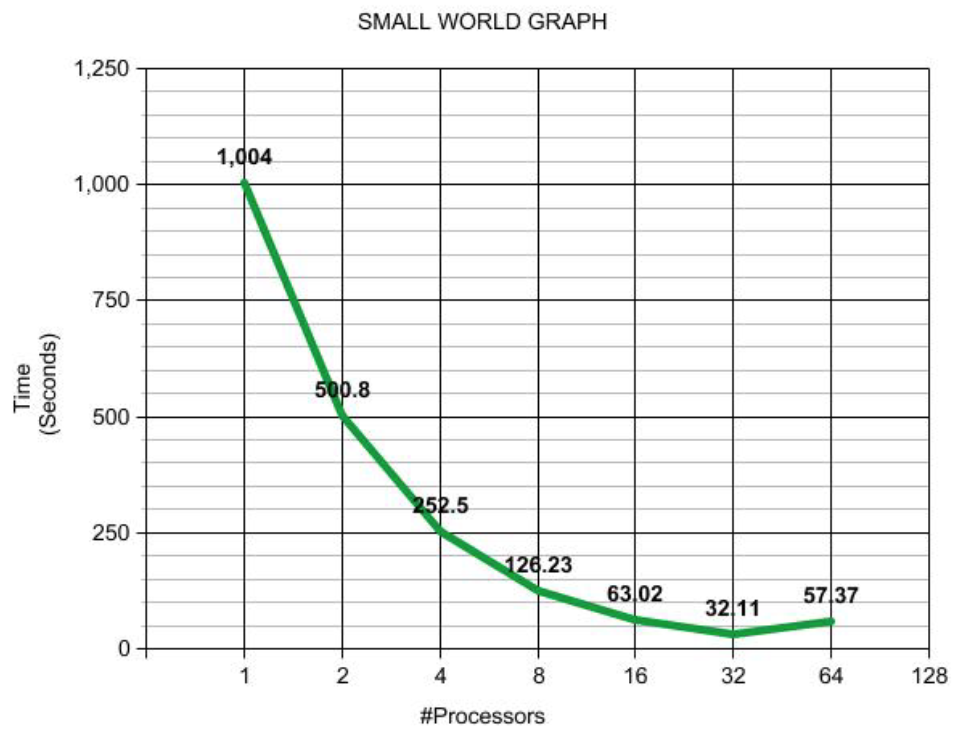}
  \caption{Scalability Results for Small world graphs}
  \Description{}
\end{figure}

Similar results we got for small world graphs as it showed
very nice scalability till 32 processors. After 32 processors
when we checked it for 64 processors it again increased which
is caused by domination of message passing overhead between
processors over computation time as was in the case of Erdos
Renyi graphs.

\subsection{Issues faced with testing environment}
\begin{enumerate}
\item Initially we were unable to generate 4 million nodes on CCR (our testing environment) due to out of memory error stated as
bad\_alloc on our test runs.
\item CCR took 3-4 days to bring our jobs from pending state to
running state and if any memory error occurred which did in
our case as mentioned above, we got to know about it after
these many days which resulted in testing of only star graphs
for 4 million vertices and other graphs for less number of
vertices due to time constraint.
\item Time limit on job execution imposed by CCR is 72 hours
but in case of our execution of 4 million nodes on 1, 2 and 4
processors took more than that it resulted in job discarded by
CCR.
\end{enumerate}

\section{Optimizations}
We made two optimizations on the algorithm in \cite{bulucc2011parallel} in further reducing execution time.\ These optimizations are:

\subsection{Algorithm optimizations}
\begin{enumerate}
\item Algorithm in described paper involves adding current vertex
to local buffer which will sent to the owner processor in the
communication step. However, we added conditional check to
see if current processor is owner of neighbor vertex and
updated distance vector in the very same step. This
optimization resulted into relatively lower buffer size and will
prove effective over large data communication.
\item Our optimization lies in the very step of all to all
communication step where algorithm in paper \cite{bulucc2011parallel}, all the local
buffers are aggregated into one local buffer each processor
which are then scattered among different processors. We were
able to send local buffers to other processors directly and
saved communication overhead of sending, merging and again
scattering. Processors are able to update distance vectors after
above communication optimization step. Such optimization
will enhance the working time of algorithm and
communication overhead which increases linearly with the
number of processors can be saved. So these were two major
optimizations which reduces execution time furthermore and
helped us achieving better results.
\end{enumerate}

\subsection{Programming optimizations}
Our program implementation consists of the following steps:
\begin{enumerate}
\item When all processors after traversing through neighbors of
vertices send populated local buffers to other processors so
that each processor should get vertices of which it is the
owner.
\item At end of algorithm when we were merging all NS vectors
to form new FS vector at master for next iteration.
Firstly we were using MPI\_Send and MPI\_Recv which were
taking 32 seconds when we ran for 10,000 vertices then we
replaced it with MPI\_Gatherv at both steps, same run with
same input and number of processor resulted in 11 seconds of
execution time.
\end{enumerate}

\section{Conclusions and Future Work}
With the results achieved in graphs in previous section the
fundamental understanding that we now have of scalability
properties of algorithms is perhaps as important as the
software implementation and results obtained.
As our results has shown that it is possible to achieve a
scalable solution for challenging BFS graph problems on
distributed memory architecture, the next big challenge would
be in developing softwares which are portable across different architectures. But still after the results we felt somewhere we
can apply number of tweaks to algorithm to make it much
more efficient in terms of memory usage, execution time and
many other factors. For example, by using some other method
of partitioning vertices it might be possible to get much more
efficient vertices distribution among processors. Also by
inventing some other methods of graph production like by
reading it from file it might be possible to make processors
from from graph production and in memory graph storage.
Which could further free CPU for more computations and
might give further reduces execution times.
In this limited period of time we tried our level best to
produce as optimal algorithm as possible but surely in future
we will apply other tricks and tweaks to algorithm which
could help reduce its execution time further.

%
\bibliographystyle{ACM-Reference-Format}
\bibliography{sample-base}

%

\end{document}